\documentclass[amssymb,amsmath,amsthm,superscriptaddress,
reprint, aip, cha]{revtex4-2}
\usepackage{lmodern}
\usepackage{mathpazo}
\usepackage{graphicx}	
\usepackage{dcolumn}	
\usepackage{bm,bbm}		
\usepackage{upgreek}
\usepackage{xcolor}		 

\usepackage[bookmarks=false]{hyperref}	
\hypersetup{colorlinks=true,citecolor=blue,linkcolor=red,
	urlcolor=blue,pdfstartview=FitH,bookmarksopen=true}
\makeatother

\usepackage{csquotes}
\MakeOuterQuote{"}

\usepackage{soul}	
\soulregister\cite7
\soulregister\ref7
\soulregister\citep7
\soulregister\citet7
\soulregister\pageref7

\newcommand{\rev}{\textcolor{black}}

\newcommand{\kb}[2]{\vert #1 \rangle \langle #2 \vert}

\newcommand{\kt}[1]{\vert #1 \rangle}

\newcommand{\ev}[1]{\langle #1 \rangle}

\newcommand{\AffiBAQIS}{\affiliation{Beijing Academy of Quantum Information Sciences, Beijing 100193, China}}
\newcommand{\AffiBUPT}{\affiliation{School of Science and State Key Laboratory of Information Photonics and Optical Communications, Beijing University of Posts and Telecommunications, Beijing 100876, China}}

\begin{document}
	
	\title{Steering nonlocality in high-speed telecommunication system without detection loophole}

    \author{Qiang Zeng}	
    \email{zengqiang@baqis.ac.cn}
    \AffiBAQIS
    \author{Huihong Yuan} 	
    \thanks{These authors contributed equally to this work.}
    \AffiBAQIS
	\author{Haoyang Wang} 	\AffiBAQIS \AffiBUPT
	\author{Lai Zhou} 		\AffiBAQIS
	\author{Zhiliang Yuan}	\email{yuanzl@baqis.ac.cn}	\AffiBAQIS
	
	\date{\today}
	
\begin{abstract}
    Nonlocal correlation represents the key feature of quantum mechanics, and is an exploitable resource in quantum information processing. 
    However, the loophole issues and the associated applicability compromises hamper the practical applications.
    We report the first \rev{time-bin entangled} detection-loophole-free steering nonlocality demonstration in a fully chip-fiber telecommunication system, with an ultra-fast measurement switching rate (1.25~GHz).
    In this endeavor, we propose the phase-encoding measurement scheme to adapt the system to time-bin degree of freedom, and design and fabricate a low-loss silicon chip for efficient entanglement generation. 
    An asymmetric configuration is introduced to mimic the active measurement implementation at the steering party thus bypassing the phase modulation loss.
    Consequently, we build a fiber-optic setup that can overcome the detection efficiency required by conclusive quantum steering with multiple actively switched measurement settings. 
    Our setup presents an immediate platform for exploring applications based on steering nonlocality, especially for quantum communication.
\end{abstract}
	
\maketitle
	
\section*{Introduction}\label{Sec1}
	Quantum steering, a newly defined form of nonclassical correlation~\cite{wiseman2007Steering}, has attracted considerable attention in recent years~\cite{uola2020Quantum,xiang2022QuantumSteering}.
	By nonclassical, one refers to the fact that the correlation function can not be fully reproduced by models involving local hidden variables (LHVs)---even at an 
    astronomical separation~\cite{hensen2015LoopholefreeBell,shalm2015Strong,giustina2015SignificantLoopholeFree,storz2023LoopholefreeBell}.
	The most famous form of nonclassical correlation is Bell nonlocality~\cite{brunner2014Bell}, in which no assumption is made on the outputs of the tested objects, namely, the outputs are directly determined by LHVs---this signifies the ultimate form of locality, and the disapproval of which thus indicates the strongest nonlocality.
	When one assumes that 
    the tested objects are quantum states 
    with the outputs complying with the laws of quantum mechanics (QM), the LHVs can determine only which of the states are tested, and hence the exclusion of LHVs proves only the objects as a whole is nonseparable (or quantum entanglement~\cite{horodeckiQuantumEntanglement2009} in a more popular way). 
	When considering the case where just part of the objects are confined by QM, we have the definition of quantum steering, in which the beyond-QM part is often called the "steering party" and the QM part the "steered party"~\cite{jones2007Entanglement}.
	
	With rigorous formulation and proof, 
    quantum steering fits firmly between Bell nonlocality and quantum entanglement~\cite{saunders2010Experimental}.
    It has gone beyond just a theoretical concept~\cite{reid1989Demonstration,cavalcanti2009Experimental,he2013GenuineMultipartite,kogias2015Quantification,degois2023CompleteHierarchy} and has been demonstrated in continuous-variable systems~\cite{handchen2012Observation,deng2017DemonstrationMonogamy,wang2020DeterministicDistribution,deng2021SuddenDeath,liu2022DistillationGaussian}, discrete systems of two dimensions~\cite{zeng2020reliable,tischler2018ConclusiveExperimental,kocsis2015Experimental} or higher~\cite{zeng2018,quRetrievingHighDimensionalQuantum2022,srivastavQuickQuantumSteering2022}, and of various degrees of freedom (DOFs)~\cite{slussarenkoQuantumSteeringVector2022,armstrong2015MultipartiteEinstein,weston2018Heralded,zhao2020ExperimentalDemonstration,zhaoDeviceindependentVerificationEinstein2023}.
    Similarly to Bell nonlocality and quantum entanglement, quantum steering is not only of fundamental interest but also has practical importance. 
    For example, it has found various applications, such as one-sided device-independent quantum secure communication (1sDI-QKD)~\cite{branciard2012Onesided}, secure quantum teleportation~\cite{reid2013Signifying}, secret sharing~\cite{xiang2017Multipartite}, randomness certification~\cite{skrzypczyk2018Maximal,guo2019Experimental}, and subchannel discrimination~\cite{piani2009All}.

	Certifying steering nonlocality with physical apparatuses could be spoiled by security loopholes, such as locality~\cite{pironio2009DeviceindependentQuantum}, freedom-of-choice\cite{wittmann2012Loopholefree}, and most notoriously detection~\cite{gisin1999LocalHidden} loopholes closing which often demands the system with low noise and high collection efficiency.
	The event-ready implementation of steering could bypass the efficiency barrier yet requires the quantum repeater~\cite{weston2018Heralded}.
	To overcome the efficiency barrier, most existing approaches change the measurement settings manually in order to avoid losses arising from active switching, which not only will open the locality loophole but more importantly hinders quantum steering from practical applications~\cite{bennet2012Arbitrarily,smith2012Conclusive,slussarenkoQuantumSteeringVector2022}.
	To date, the only loophole-free experiment~\cite{wittmann2012Loopholefree} achieves a switching rate up to 0.787~MHz using polarization DOF, which employs the three-setting Platonic-solid measurement scheme~\cite{saunders2010Experimental}.
	Such a measurement scheme has the measurement bases at the steering party uniformly distributed on the Bloch sphere.
	Though the scheme has achieved the optimal performance in detecting quantum steering, \rev{it requires amplitude modulation between the orthogonal bases if requiring more measurement settings (larger than three).
	Note that amplitude modulation could introduce cross-talk and extra loss in practical implementation.}
	Moreover, all existing experiments rely on the free-space configuration (at least at the entanglement-generation phase) to preserve the necessary detection efficiency, but at the cost of system robustness and practicality.
	So far, a conclusive demonstration of steering nonlocality without applicability compromises is still lacking.
	
	\rev{In this work, we provide an alternate measurement scheme which employs no amplitude but purely phase modulation to accommodate steering test to time-bin DOF and particularly telecommunication hardware thus achieving a GHz-level switching rate.}
	We further quantitatively compare the performance of Platonic-solid and our phase-encoding measurement schemes, and find that they are almost the same when the noise is sufficiently low.
	Experimentally, we validate our scheme and certify steering nonlocality in a fully chip-fiber-based manner with the detection loophole closed. 
	To overcome the high loss of phase modulation devices, we made a low-loss silicon chip and came up with a method of translating the phase modulation from measurement to before entanglement generation to mimic the active measurement implementation at the steering party.
 	Our work for the first time brings detection-loophole-free steering nonlocality certification to a fully chip-fiber-based setup, notably with a 1.25~GHz measurement switching rate at the steered side.
 	The system we built is versatile and field-use-friendly for it is easy to integrate, adapting to high-speed telecommunication modules, and robust to noise, thereby representing a solid step of steering nonlocality research from laboratory to application.
	
\section*{Results}\label{Sec2}
\subsection*{Steering test}\label{}
	Here we consider the bipartite qubit scenario, in a steering test, the examiner (Bob) first receives a set of quantum states (called assemblage), and then selects a list of questions from a questionary to send them to the examinee (Alice).
	Each question has binary answers.
	Bob obtains his answer bit from the measurement outcome of the local quantum states prepared by Alice.
	We denote Bob's measurement by $y$ and the outcome by $b$.
	In the meantime, Bob makes no assumption on the method that Alice would take to get her answer bit, which we denote by $A$.
	After sufficient runs of questioning, Bob combines both of their answers to check the scores (correlation).
	Note that since the local quantum states are prepared by Alice, she can predict with certainty the outcome Bob would get if a specific question (measurement) is raised.
	However, this preset strategy has a limit in terms of correlation in the locality frame, similar to the classical bound for Bell inequality~\cite{bell1964Einstein}.
	Formally, one terms the preset quantum states \emph{local hidden state model} (LHSM) and defines it as
	\begin{equation}\label{LHS}
		\sigma^{\text{LHS}}_{b|y}=\int \sigma_{\lambda}q(\lambda) p(b|y,\lambda)d\lambda,~\forall b,y,
	\end{equation}
	where $\{\sigma_\lambda\}$ is a set of positive matrices with some probability distribution $q(\lambda)p(b|y,\lambda)$, and $\lambda$ is the local hidden variable held by Alice.
	Accordingly, one can define the steering parameter~\cite{evans2013Losstolerant}
	\begin{equation}
		S_n \equiv \frac{1}{n}\sum_{k=0}^{n-1}\langle A_k \hat\sigma^B_k \rangle,
	\end{equation} 
	which quantifies the correlation between the two parties, where $n$ is the number of measurements, $A_k \in \{-1,1\}$, $k \in \{0,...,n-1\}$, and $\hat\sigma^B_k$ is the measurement chosen by Bob with $\ev{\hat\sigma^B_k} \in \{-1,1\}$.
    \rev{It is worth mentioning that the roles of the two parties can be swapped, and a similar steering parameter applies.}

	Without considering loopholes, it is safe to say that locality is breached by Alice if the answers she provides are strongly correlated thus surpassing the limit~\cite{evans2014Optimal}.
	However, if there exists a hidden signaling between the parties, or Alice somehow can influence the measurement choice of Bob, she can easily fake a perfect correlation, which in fact opens the locality and free-of-choice loophole, respectively.
	In a trickier scenario, Alice may ignore the question Bob raised if the question is not the specific one she prepared.
	By reporting a null answer, Alice can significantly improve the correlation result, and there is in principle no way to verify whether the questions are fairly sampled.
	This leads to the fair-sampling assumption, which is known also as the detection loophole from experimental perspectives.
	
	\rev{The locality and free-of-choice loophole are usually dealt with by enforcing a space-like separation between participants and using trusted random number generators.}
	To address the detection loophole, one needs to consider a loss-tolerant LHSM to fairly evaluate the answers Alice reports. 
	For the construction of such LHSMs, it is crucial to carefully choose Bob's measurement settings.
	In this regard, the optimal set of measurements for detecting steering nonlocality using two-qubit Werner states has been investigated~\cite{evans2014Optimal}, in the sense that it requires the minimal entanglement resource to demonstrate steering nonlocality given certain detection efficiency.
	However, this set of measurements is not compatible with the telecommunication system for it involves amplitude modulation of bases.
	
	A recent study shows that it is possible to upper-bound the maximal correlation an LHSM could reach given any set of input settings~\cite{zeng2022OnewayEinsteinPodolskyRosen}.
	In specific, this method reformulates the LHSM in the form of semidefinite
	programming (SDP)~\cite{skrzypczyk2014Quantifying,cavalcanti2017Quantum}, and considers the discarded outcomes arising from a mixed state that encompasses all the possible local states Bob might hold.
	As the mixed state contributes zero nonlocal correlation to the result, the effective correlation of the LHSM from all conclusive detection events can be extracted~\cite{skrzypczyk2015Losstolerant}.

\begin{figure}[t]
	\includegraphics[width=.85\columnwidth]{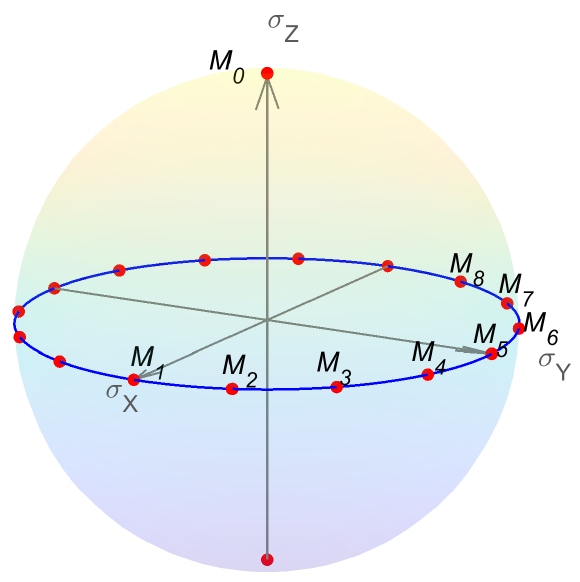}
	\centering
	\caption{\label{F1} 
			\textbf{The diagram of Bob's measurements.}
			In the Bloch sphere representation, $M_0$ corresponds to $\hat\sigma_{\textsf{Z}}$, and the rest of the measurements are uniformly distributed on the equator.
	}
\end{figure}

\begin{figure*}[t]
	\includegraphics[width=2\columnwidth]{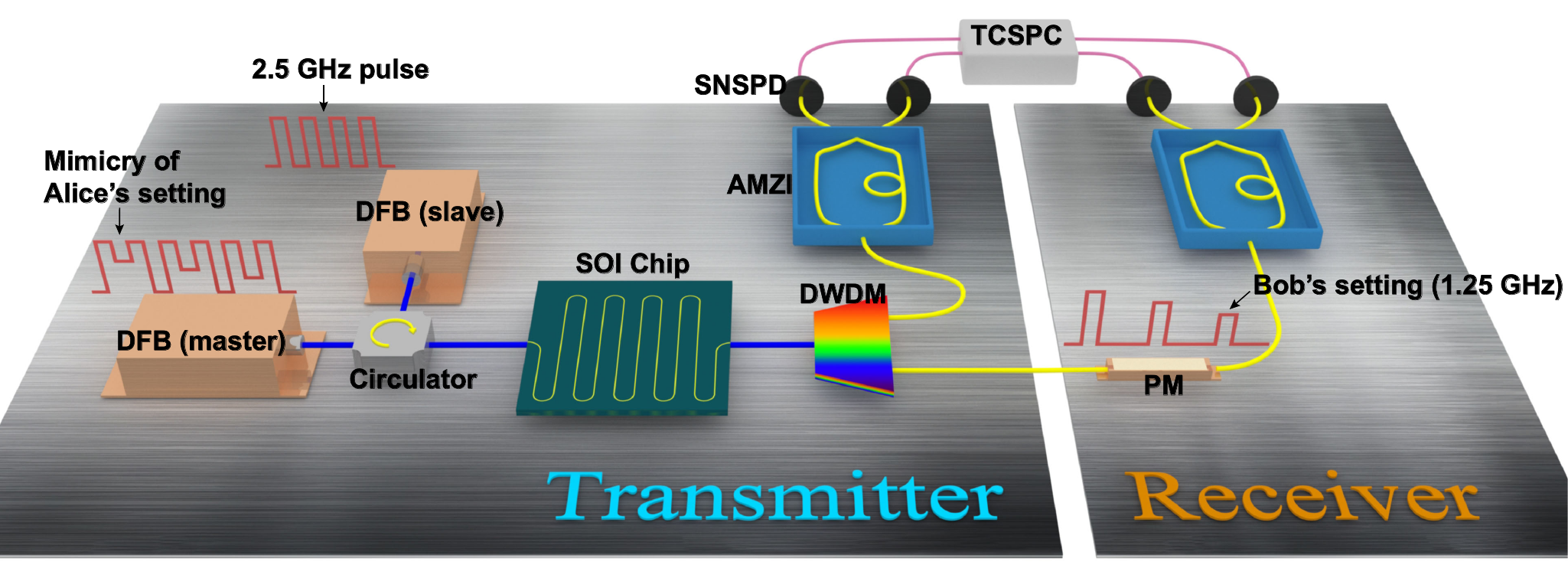}
	\centering
	\caption{\label{F2}  
			\textbf{Experimental setup.}
			The setting signals on both sides and the 2.5~GHz pulse signal on the slave DFB are generated by an arbitrary wave generator (AWG) (not shown in the figure).
				PM: phase modulator; DFB: distributed feedback laser; DWDM: dense wavelength division multiplexer; AMZI: asymmetric Mach-Zehnder interferometer; TCSPC: time-correlated single photon counting; SNSPD: superconducting nanowire single-photon detector.
	}
\end{figure*}

\subsection*{Phase-encoding method}\label{Sec3}
    Time basis is commonly employed to encode information in telecommunication systems for its resistance to decoherence.
	\rev{Coherently splitting a single photon equally into two time-bins, so that it can either be detected at an early moment $t_0$, or at a late moment $t_1$.
	If we denote the state at $t_0$ as $\kt{\text{E}}$, and the state at $t_1$ as $\kt{\text{L}}$, a qubit can be defined in the time DOF:
	$\kt{\psi}=(\kt{\text{E}}+e^{\text{i} \phi_0}\kt{\text{L}})/\sqrt{2}$.}
	Note that the superposition of the two states is limited by the coherence time of the photon.
	
	By recording the arrival time, one essentially performs $\hat\sigma_{\textsf{Z}}$ measurement on the photon, producing one of the two temporal outcomes.
    Further, by exploiting an asymmetric Mach–Zehnder interferometer (AMZI) which delays the early state to temporally overlapping with the late state, one obtains a different set of two incompatible outcomes from the two outputs of the interferometer and thus allows inference of retrieving the phase information. 
    This corresponds to a $\hat\sigma_{\theta}$ measurement
	\begin{equation}
		\hat\sigma_{\theta}=\cos(\theta)\hat\sigma_{\textsf{X}}+\sin(\theta)\hat\sigma_{\textsf{Y}},
		\nonumber
	\end{equation}
	where $\theta$ is the differential phase delay of the AMZI and $\hat\sigma_{\textsf{X,Y,Z}}$ refers to the Pauli operators.

	The phase modulation of $\theta$ 
 	can be accurately and swiftly implemented by using an electro-optical modulator.
	Thus it is convenient for Bob to construct an $n$-number measurement set $\{M_i\}_n$, with $M_0=\hat\sigma_{\textsf{Z}}$ and $M_j=\hat\sigma_{\frac{j-1}{n-1}\pi},~j \in \{1,...,n-1\}$.
	\rev{Figure~\ref{F1} gives the example of the measurement set when $n=9$ in the Bloch sphere representation, in which the measurement set forms a regular right symmetric $m$-gonal bipyramid with $m=2(n-1)$.}
	
	Using the numerical methods~\cite{zeng2022OnewayEinsteinPodolskyRosen}, we evaluate the performance of the phase-encoding measurement scheme introduced above, and find it is suboptimal when compared with the conventional Platonic-solid measurement scheme~\cite{evans2014Optimal} in terms of quantifying the steering nonlocality for a given state.
	However, with only phase-related measurement settings plus one computational basis setting, our scheme is compatible with the high-speed telecommunication system.
	We also find that when the noise is low, the two schemes have fairly close performance.
	An elaborate comparison between the phase-encoding measurements and the Platonic-solid measurements is provided in Supplementary Note 1.

\subsection*{Experimental setup}\label{Sec5}
	In a typical steering test, the steering party (Alice) who is untrusted gets access to the measurement setting only after the steered party (Bob) performs his measurement or receives the quantum state to be measured.
	Practically, this causal requirement leads to the symmetrical configuration in which both Alice and Bob place modulation devices at their measurements.
	However, the modulation introduces extra losses at the steering side.
	In specific, a phase modulator typically introduces about 3~dB insertion loss.
	\rev{To bypass the loss, we develop an \textit{ad hoc} configuration to translate the modulation of Alice from measurement to entanglement generation. 
	We note that this asymmetric configuration is for proof-of-principle demonstration with the locality loophole implied (see Supplementary Note 2 for discussion), and the symmetric configuration is needed when it comes to application scenarios.}
        
	Our asymmetric configuration proceeds as follows.
	Firstly, instead of a fixed entanglement state, we consider a set of two-qubit isotropic states entangled in time-bin DOF
	\begin{equation}\label{iso}
		\rho_{AB}=\left\{v\kb{\Psi(\alpha)}{\Psi(\alpha)}+(1-v)\mathbbm{I}/4,~0 \le \alpha \le \pi \right\},
	\end{equation}
	\rev{where $\kt{\Psi(\alpha)}=\left(\kt{\text{E}}_A\kt{\text{E}}_B+e^{\text{i}\alpha}\kt{\text{L}}_A\kt{\text{L}}_B\right)/\sqrt{2}$}, and $v$ denotes the visibility of the maximally entangled state.
	We assume that the experimental noises and imperfections are reflected by the isotropic noise $\mathbbm{I}/4$ that hampers the visibility, with a coefficient $1-v$, which is reasonable because the erroneous correlation data is unbiased to measurement settings (see Supplementary Note 6 for discussion).
			
	Secondly, we apply the phase modulation (the $\alpha$ in $\kt{\Psi(\alpha)}$) based on injection-locking technique~\cite{yuan2016Directly} prior to the entanglement generation process, and regard this modulation as the mimicry of measurement implementation of Alice.
	\rev{The actual measurement at Alice's side is fixed, and is regarded as the consequence of Bob's choice.}
	A detailed explanation on the equivalence of our asymmetric configuration and the convectional symmetric configuration in terms of revealing the nonlocal correlation of the quantum system is provided in Supplementary Note 2.
    \rev{It is important not to confuse the aforementioned asymmetric configuration with the intrinsic asymmetry of steering nonlocality~\cite{zeng2022OnewayEinsteinPodolskyRosen}.
    The former describes the experimental setup which aims to close the detection loophole at the steering side, while the latter describes the (potential) asymmetry of the entanglement state.
    In particular, the state we adopt in Eq.~\eqref{iso} is symmetric, and we demonstrate steering from Alice to Bob in accordance with our asymmetric configuration.
    }

 \begin{figure}[t]
 	\includegraphics[width=\columnwidth]{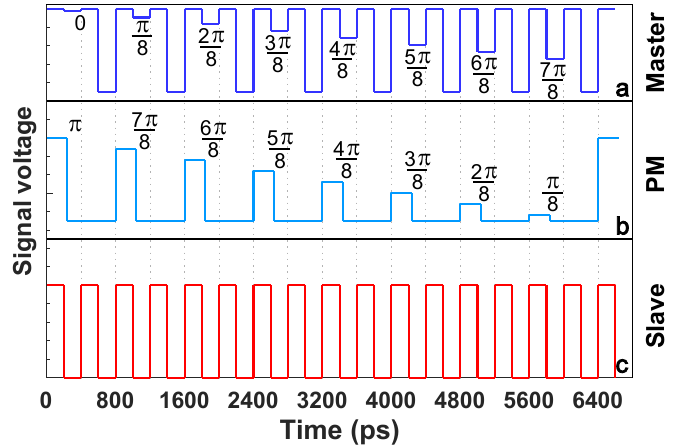}
 	\caption{\label{F3} 
 			\textbf{Schematic of signal voltages.} 
 			Taking $n=9$ as an example, from top to bottom: 
 			(a) driving signal in the master DFB based on injection-locking technique; 
 			(b) driving signal in the phase modulator; 
 			(c) 2.5~GHz square wave signal in the slave DFB.
 			The master laser and PM introduce complementary phases, which is equivalent to Alice and Bob performing complementary measurements.
 	}
 \end{figure}
	
\begin{figure*}[htp]
	\includegraphics[width=2\columnwidth]{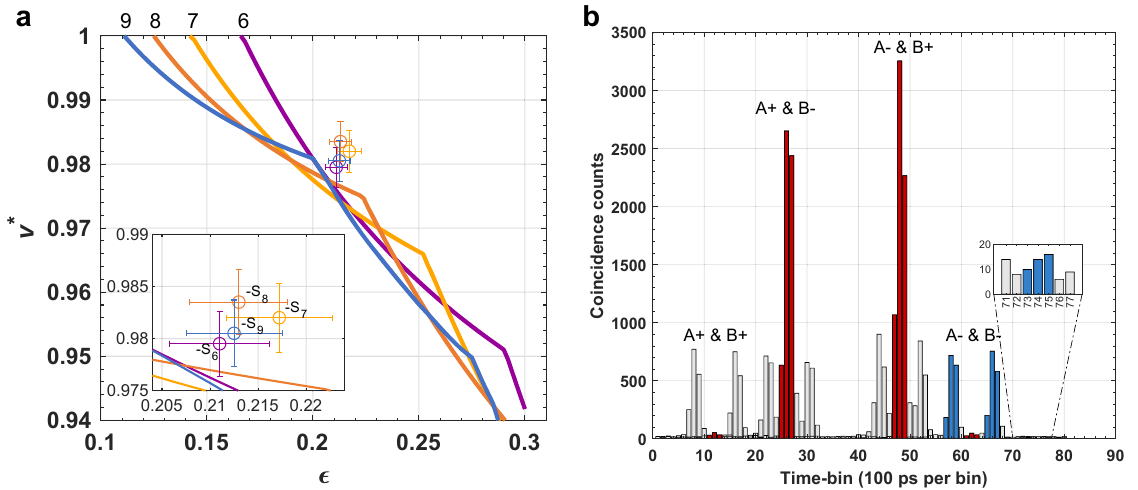}
	\caption{\label{F4} 
			\textbf{The results of steering test.} 
			\textbf{a.} The colored circles denote the measured steering parameters $-S_n$ which quantify the degree of correlation.
			The bound of critical $v^*$ shown by solid lines is computed based on the loss-tolerant LHSM for $n=6,7,8,9$ respectively. 
			\rev{\textbf{b.} Pairwise time-bin histograms of AMZIs. 
				The notations "A$\pm$" and "B$\pm$" denote the output ports of Alice's and Bob's AMZIs respectively.
				Correlation results in time basis are painted with blue (exemplified by A$-$ \& B$-$), and the ones in phase basis are in red.
				The erroneous results in time basis are estimated by the accidental coincidence counts in the adjacent time-bins (see the inset).}
	}
\end{figure*}

    Our experimental setup is sketched in Fig.~\ref{F2}.
    We take advantage of the spontaneous four-wave mxing (SFWM) effect in a silicon-on-insulator (SOI) spiral waveguide chip to generate the required quantum states, \rev{and the on-chip photon-pair generation rate (PGR) achieves $1.3\times10^6$~pairs~s$^{-1}$~mW$^{-2}$.}
    The design and fabrication details about the SOI chip \rev{and the experimental data of PGR} are provided in Supplementary Note 3.
	The pump light is generated through a pair of distributed feedback (DFB) lasers concatenated via a circulator.
	The slave laser is gain-switched by square waves to produce 2.5~GHz light pulses, and the relative phases of these pulses are modulated by rapidly adjusting the output power of the master laser~\cite{zeng2023ControlledEntanglement}.
	By impacting the modulated pump light into the SOI chip, we produce the set of states in Eq.~\eqref{iso}.
	Passing through the SOI chip, two pump photons probabilistically annihilate for the creation of two entangled daughter photons. 
	This third-order nonlinear process conserves energy and momentum, resulting in a broadband signal-idler spectrum.
	A dense wavelength division multiplexer (DWDM) module is then used to divide the spectrum into multiple wavelength channels.
	Here, we choose to use channels C29 and C38 (ITU grid specification; Full width at half maximum (FWHM): 100~GHz) for our demonstration.
	We also insert a filtering module with FWHM of 10~GHz after Bob's AMZI (not shown in the figure) to overcome the filtering inefficiency~\cite{meyer-scott2017LimitsHeralding}.
		
	The idler photon is directed into a homemade AMZI before local detection at Alice's site.
	On the other hand, the signal photon is sent over a single-mode fiber to the receiver Bob. 
	After the modulation by the phase modulator (PM), which corresponds to Bob implementing measurements, the signal photon then feeds through Bob's AMZI and is sent to the detector.
    Alice and Bob's AMZI's have a nominally identical differential delay of 400~ps. 
    They are independently phase-locked through temperature feedback on the interference of a continuous-wave reference laser that has the same central wavelength as Alice's master DFB laser.
    For a detailed description on phase locking, see Supplementary Note 4.

	To demonstrate steering nonlocality with the above setup, we consider the phase-encoding measurement set on Bob's side, i.e., $\{M_i\}_n$, with $M_0=\hat\sigma_{\textsf{Z}}$ and $M_j=\hat\sigma_{\frac{j-1}{n-1}\pi}$.
	We perform the complementary measurement $\{\overline{M_i}\}_n$, with $\overline{M_0}=-\hat\sigma_{\textsf{Z}}$ and $\overline{M_j}=\hat\sigma_{\left(1-\frac{j-1}{n-1}\right)\pi}$ on Alice's side, though we do not assume the measurement is faithfully executed.
	The illustration of how the phase-related operators of $\{M_i\}_n$ and $\{\overline{M_i}\}_n$ correspond to the signal voltage respectively on the PM and the master laser is depicted in Fig.~\ref{F3}, where the signal voltage of the slave laser works as a temporal reference \rev{(See Supplementary Note 5 for technical details).}

\subsection*{Test results}\label{Sec6}
	Should Alice indeed send half of the entangled pair to Bob, we can verify that $S_n=-v$, noting that $\kt{\Psi(\alpha)}$ is perfectly anti-correlated with $\{M_i\}_n$ and $\{\overline{M_i}\}_n$ being measured, and $\mathbbm{I}/4$ has no contribution to $S_n$.
	\rev{On the other hand, if Alice exploits the detection loophole and constructs an LHSM, she can reproduce the correlation results of the entanglement state.}
    \rev{To demonstrate that the detection loophole is closed, a large steering parameter surpassing the criterion is required meaning that even with the help of detection loophole, the LHSMs fail to fully reproduce the observed correlation, which in turn verifies the detection-loophole-free nonlocality.}
    
	We denote by $v^*$ the critical visibility that a detection-loophole-assisted strategy could ever achieve, given certain detection efficiency $\epsilon$.
	As we mentioned before, the relation between $v^*$ and $\epsilon$ can be formulated in a semidefinite programming problem, and can be numerically computed~\cite{zeng2022OnewayEinsteinPodolskyRosen}.
	In Fig.~\ref{F4}\textbf{a}, we plot with solid lines the relation of $v^*$ -- $\epsilon$ for $n=6,7,8,9$ respectively.
	\rev{It shows that the benefit of increasing measurement settings is to tolerate a lower critical detection efficiency that permits closure of the detection loophole given certain $v$, when $v$ is approaching unity.}
	

    To conclusively certify steering nonlocality, it thus requires $-S_n>v^*$.
	It is worth noting that a full state tomography to estimate $v$ is unnecessary, because (i) certifying steering nonlocality does not rely on the assumption that the state is entangled, (ii) drawing conclusions based on high fidelity can be inaccurate~\cite{peters2004MixedstateSensitivity,tischler2018ConclusiveExperimental,zeng2020reliable}.
	The remaining question lies in how to estimate $S_n$ from the time-recording data.
	Specifically, for each pair of outputs of the AMZIs, there are three possible photon arrival times, which results in three equidistant coincidence peaks in the arrival-time histogram. 
	The coincident events in the side peaks yield detections in the time basis ($\hat\sigma_{\textsf{Z}}$), according to which we compute the term of $k=0$ in $S_n$.
	The coincidence which corresponds to the uncorrelated results in time basis is estimated by adopting the counts in the adjacent time-bins.
	Although the time basis is passively selected depending on the path that the photon takes at the first beam-splitter of the AMZI with a probability of 50\%, this probabilistic process does not introduce postselection loophole~\cite{vedovatoPostselectionLoopholeFreeBellViolation2018} as the data collected in time-basis contributes equally to the nonlocality analysis.
	On the other hand, the coincident events in the central peak yield detections in the phase basis ($\hat\sigma_{\theta}$), according to which we compute the rest of the terms in $S_n$.
	\rev{An example of our time-recording data depicting by pairwise time-bin histograms is presented in Fig.~\ref{F4}\textbf{b}. }
	The detailed method to obtain $S_n$ are elaborated in Supplementary Note 6.

	\rev{The measured losses on Alice's side include on-chip transmission ($-2$~dB), chip-fiber coupling ($-1.5$~dB), DWDM ($-1$~dB), fiber coupling ($-0.1$~dB), AMZI ($-1$~dB), filtering inefficiency ($-0.5$~dB), and SNSPD ($-0.5$~dB), and the overall loss corresponds to the collection efficiency ($-6.6$~dB). 
    In the detection loophole context, we attribute all losses to the detector as one in principle cannot identify where exactly the specific photon is lost.}
	Here, we adopt the Klyshko efficiency~\cite{klyshkoUseTwophotonLight1980a} (also known as heralding efficiency, defined as the coincidence counts divided by the single counts of the opposite arm) to estimate the detection efficiency.

	The results of $S_n$ and $\epsilon_n$ are presented in Fig.~\ref{F4}\textbf{a}.
	In specific, $S_6=-0.9795\pm0.0031$, $S_7=-0.982\pm0.0033$, $S_8=-0.9835\pm0.0031$, $S_9=-0.9805\pm0.0032$, and $\epsilon_6=0.211\pm0.0052$, $\epsilon_7=0.2172\pm0.0055$, $\epsilon_8=0.213\pm0.0051$, $\epsilon_9=0.2125\pm0.005$.
	It is clear that given certain detection efficiency, the obtained $-S_n$ for $n=6,7,8,9$ lie above the critical $v^*$, which indicates that the steering test is passed, and notably, without detection loophole.

\section*{Discussion and outlook}\label{Sec8}
	Although the theory allows the critical detection efficiency down to $11.1\%$ with $n=9$, the experimental noises and imperfections hamper the system from achieving the ideal performance.
	We attribute the non-unit degree of correlation in the phase basis mainly to the imperfect interference of the AMZIs, whose interference visibility is about $99.2\%$ each.
	Regarding the erroneous correlation results in the time basis, we attribute it to the dark counts of the SNSPD which is about 100~Hz and the multi-photon components of the pump light which becomes significant when the pump power increases.
	When it comes to field tests, another issue that may significantly impact the correlation results is chromatic dispersion, as the transmitter and the receiver could be well separated, and in this case the dispersion compensation modules are required.
	
	One promising application of our platform is the well-known 1sDI-QKD protocol~\cite{branciard2012Onesided}.
	For two measurement settings, the critical $\epsilon$ for 1sDI-QKD is about 65.9\%~\cite{branciard2012Onesided}, which is beyond the performance of our present setup.
	However, as proved in Ref.~\cite{branciard2012Onesided}, securing 1sDI-QKD protocol amounts to demonstrating quantum steering, and thus we note that the critical $\epsilon$ for a multi-setting 1sDI-QKD protocol can be significantly reduced through increasing the number of measurement settings.
	As in our case the loophole at the transmitter side is closed, we propose that a transmitter-device-independent QKD (TDI-QKD) can be realized based on our work.

	

\section*{DATA AVAILABILITY}
{	
	\small
	The data that support the plots within this paper and other findings of this study are available from corresponding authors upon reasonable request.
}

\section*{references}
\bibliography{Refs}

\section*{acknowledgments}	
{	
	\small
	The SDPs implemented in this work used {\sc MATLAB} and the packages {\sc CVX}~\cite{cvx,gb08}.
	Q. Z. thanks Sergei Slussarenko for helpful discussion on detection efficiency.
	This work is supported by National Natural Science Foundation of China under Grants 12105010 (Q. Z.), 62105034 (L. Z.), and 62250710162 (Z. Y.).	
}

\section*{AUTHOR CONTRIBUTIONS}	
{	
	\small
	Q.Z. and H.Y. contributed equally to this work.
    Q.Z. devised the experimental setup, and performed the experiment with the assistance of H.W..
	H.Y. designed and fabricated the silicon chip.
	All authors contributed to the discussion and writing of the paper.
	Z.Y. guided the project. 
}

\section*{COMPETING INTERESTS}	
{	
	\small
The authors declare no competing interests.
}



\end{document}


\title{Supplementary Information: Steering nonlocality in high-speed telecommunication system without detection loophole}
	
	\author{Qiang Zeng}		\email{zengqiang@baqis.ac.cn}	\AffiBAQIS
	\author{Huihong Yuan} 	\AffiBAQIS
	\author{Haoyang Wang} 	\AffiBAQIS \AffiBUPT
	\author{Lai Zhou} 		\AffiBAQIS
	\author{Zhiliang Yuan}	\email{yuanzl@baqis.ac.cn}		\AffiBAQIS
	
	\maketitle
	\onecolumngrid
	\appendix

\newpage	
\section*{Note 1: Comparison between phase-encoding measurement setting and Platonic-solid measurement setting}

\begin{figure}[h]
	\centering
	\includegraphics[width=400px]{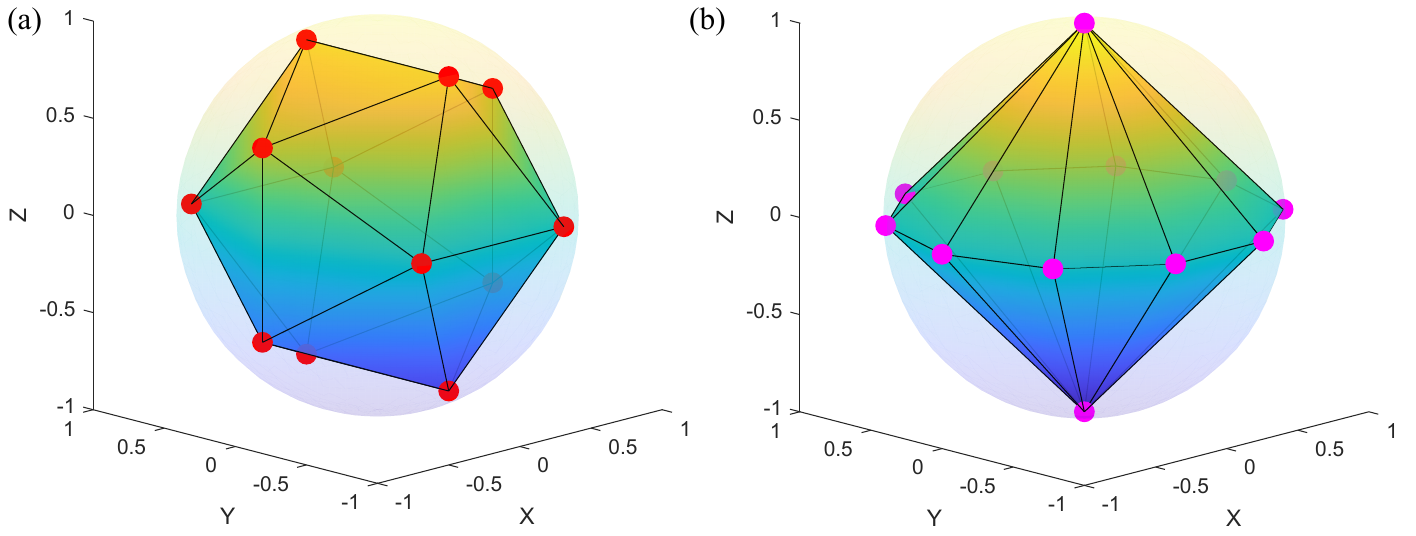}
	\caption{\label{F5} 
			\textbf{Comparison of measurement setting in Bloch sphere representation.}
			(a) Platonic-solid measurement setting.
			(b) Phase-encoding measurement setting.
	}
\end{figure}

\begin{figure}[h]
	\centering
	\includegraphics[width=240px]{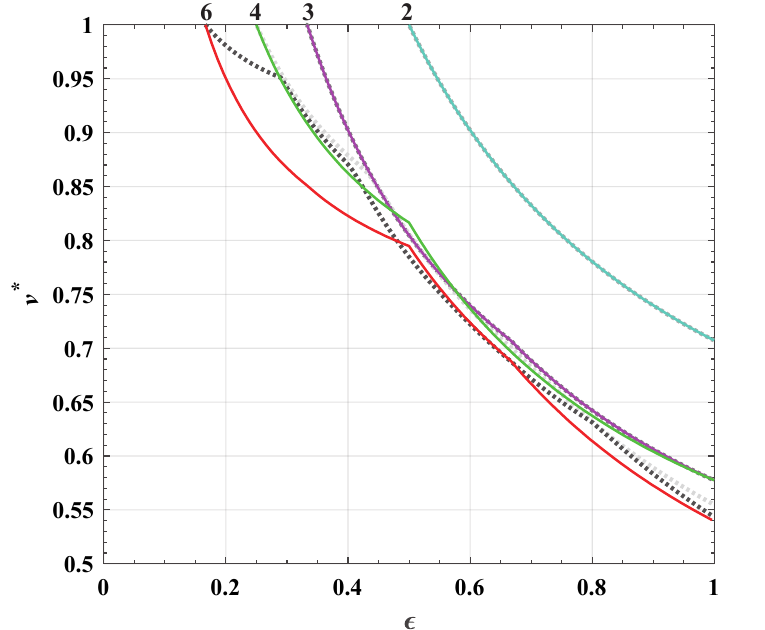}
	\caption{\label{F6}  
			\textbf{Critical bound of parameters for demonstrating steering nonlocality.}
			The solid lines denote the bound using Platonic solids setting, and the dashed lines denote the bound using phase-encoding setting.
	}
\end{figure}

	The Platonic-solid measurement setting~\cite{evans2013Losstolerant} refers to a set of measurements that correspond to the antipodal pairs of vertices of an inscribed Platonic solid within the Bloch sphere.
	For example, when the number of settings is six, it forms an icosahedron within the Bloch sphere, as shown in Fig.~\ref{F5}(a).	
	Comparably, our phase-encoding measurement setting forms a regular right symmetric decagonal bipyramid with 20 facets, as shown in Fig.~\ref{F5}(b).
	To further illustrate the capability of the two sets of setting in detecting steering nonlocality, we plot the critical bound of parameters in Fig.~\ref{F6}, with the target state being: 
	\begin{equation}
		\rho=\left\{v\kb{\Psi}{\Psi}+(1-v)\mathbbm{I}/4,~0 \le v \le 1 \right\} \nonumber,
	\end{equation}
	where $\kt{\Psi}=(\kt{00}+\kt{11})/\sqrt{2}$, and $v$ denotes the visibility of the maximally entangled state.
	Solid lines denote the results with the Platonic-solid measurement setting and dashed lines denote the results with the phase-encoding measurement setting.
	While the two settings are identical when $n=2,3$, Platonic-solid setting outperforms phase-encoding setting notably when $n=6$, in the sense that a smaller $v$ is required given certain $\epsilon$, where $\epsilon$ is the detection efficiency.
	However, it is worth mentioning that for $v$ approaching unity, which means the noise of the system is well controlled---a common requirement in practical applications, the two settings are comparable in performance.

\newpage
\section*{Note 2: The equivalence of applying Alice's measurement before/after entanglement generation in phase-encoding measurement scheme and the locality loophole}
	As the noise we are considering is isotropic, our discussion focuses only on the maximally entangled state, i.e., $\kt{\Psi(\alpha)}=\left(\kt{00}+e^{\text{i}\alpha}\kt{11}\right)/\sqrt{2}$.
	In the conventional configuration, where Alice and Bob have symmetrical measurement setup, the entangled state from the distributor is usually fixed.
	Without loss of generality, we assume that the distributed state is $\kt{\Psi(\alpha=0)}=(\kt{00}+\kt{11})/\sqrt{2}$.
	Before stepping further, it is worth mentioning that in a steering test, Bob (the steered side) trusts that quantum mechanics is correct, but doubts that whether the state he receives (a) half of an entangled pair or (b) a pure state sent by Alice (the steering side).
	If (a), then Bob receives $\kt{0}$ or $\kt{1}$ when Alice performs $\hat\sigma_{\textsf{Z}}$, or receives $\frac{1}{\sqrt[]{2}}\left(\kt{0}+e^{i\theta_A}\kt{1}\right)$ or $\frac{1}{\sqrt[]{2}}\left(\kt{0}+e^{i(\theta_A-\pi)}\kt{1}\right)$ when Alice performs $\hat\sigma_{\theta_A}$.
	Upon Alice declares her outcomes, Bob then accordingly performs $\hat\sigma_{\textsf{Z}}$ or $\hat\sigma_{\theta_B}$ on his qubit.
	For $\hat\sigma_{\textsf{Z}}$, the correlated results are straightforward.
	While for $\hat\sigma_{\theta_B}$, it is easy to verify that their results exhibits perfect correlation if $\theta_A+\theta_B=0$, noting that
	\begin{equation}\label{phasecorrelation}
	\begin{aligned}
		&\frac{1}{2}\left[
		\left(\kt{0}+e^{i\theta_A}\kt{1}\right)
		\left(\kt{0}+e^{i\theta_B}\kt{1}\right)+
		\left(\kt{0}+e^{i(\theta_A-\pi)}\kt{1}\right)
		\left(\kt{0}+e^{i(\theta_B-\pi)}\kt{1}\right)\right]\\
		=&\kt{00}+e^{i(\theta_A+\theta_B)}\kt{11}.
	\end{aligned}
	\end{equation}
	If (b), then Alice's outcomes are not expected to be obtained from the measurements.
	We make no assumption on the pre-existing strategy that Alice uses, but the correlation results should be upper-bounded by the LHSM.
	
	Now we examine what will happen in the asymmetric configuration, where Alice's setting is translated to before the entanglement generation.
	In this case, the distributed entangled states are no longer fixed, but with a phase information encoded by Alice.
	Similar to the former case, if the state is indeed entangled, then Bob receives $\kt{0}$ or $\kt{1}$ when Alice performs $\hat\sigma_{\textsf{Z}}$, or receives $\frac{1}{\sqrt[]{2}}\left(\kt{0}+e^{i(\theta_A+\alpha)}\kt{1}\right)$ or $\frac{1}{\sqrt[]{2}}\left(\kt{0}+e^{i(\theta_A-\pi+\alpha)}\kt{1}\right)$ when Alice performs $\hat\sigma_{\theta_A}$.
	\textbf{Note that $\hat\sigma_{\theta_A}$ can be fixed in this case meaning a fixed measurement setting at Alice's, which practically is equivalent to locking the phase of the AMZI at Alice's side.}
	By setting $\theta_A=0$, we finally reach identical expressions to the symmetrical case.
	It is worth noting that if one sets $\theta_A=\pi$, which means $\theta_A+\theta_B=\pi$, then the their results will produce perfect anti-correlation.
	On the other hand, if the sent states are some local pure states, then one treats them in the same way to the former case.
	With the above description, we thus prove the equivalence of modulating the phase of the maximally entangled states and modulating the phase at Alice's measurement, in terms of revealing the nonlocal correlation of quantum system.
	
	\rev{The major difference between our asymmetric configuration and traditional symmetric configuration is that \textbf{the locality loophole is inherently implied.}
	With the phase modulation translated to before the entanglement generation, Alice’s operation is always ahead of Bob’s, which seems against the causal requirement in steering test.
	However, operation is not equivalent to measurement, and Alice's actual measurements are still performed at her interferometer which practically is fixed and can be regarded as the consequence of Bob's choice, which essentially provokes locality loophole for assuming that Bob's measurement settings are not predetermined and no information exchange happens between Alice and Bob.
	In the traditional polarization bulk optics experiments, this is equivalent to fixing the polarization measurement at Alice’s, and rotating the polarization measurement at Bob’s.
	To proceed the test, one is required to vary Alice's measurement every once in a set interval of time to obtain the correlation parameter.
	Our configuration mimics a dynamic measurement shift at Alice’s instead of fixing one every once operation, but does not alter the fact that Alice is performing a passive fixed measurement.
	To be more concise, we did not realize an actual 1.25~GHz measurement switching rate on both sides, but only on the steered (Bob) side.}
	
	\rev{It is worth mentioning that leaving the locality loophole open is not problematic in the proof-of-principle context, as many works did.
	The justification, aside from the undeveloped techniques and devices in the early stage of the research, is that
	\textbf{a.} In usual Bell/steering experiments, the locality loophole can be dealt with by enforcing a space-like separation between Alice and Bob. 
	This guarantees that no sub-luminal signals (including signals mediated by some yet-unknown theory) could have traveled between Alice’s and Bob’s devices;
	\textbf{b.} In the context of DIQKD, it is sufficient to guarantee that no quantum signals (e.g. no photon) can travel from Alice to Bob. 
	This can be enforced by a proper isolation of Alice’s and Bob’s locations, and without which cryptography would not make any sense. 
	For a more detailed discussion on the loophole issues, please refer to Ref.~\cite{pironio2009DeviceindependentQuantum}.}
	
	\rev{We note that to close the locality loophole in a practical context, say secure communication, the symmetric configuration is needed, and the loss gap induced by the phase modulation can be balanced out through: 
	\textbf{a.} improving the fabrication technology to lower the transmission and coupling loss of the chip;
	\textbf{b.} integrating filtering module to the chip;
	\textbf{c.} improving the filtering design to overcome the filtering inefficiency,
	which are all technically feasible.}
	

\newpage
\section*{Note 3: SOI chip}
\begin{figure}[h]
	\centering
	\includegraphics[width=500px]{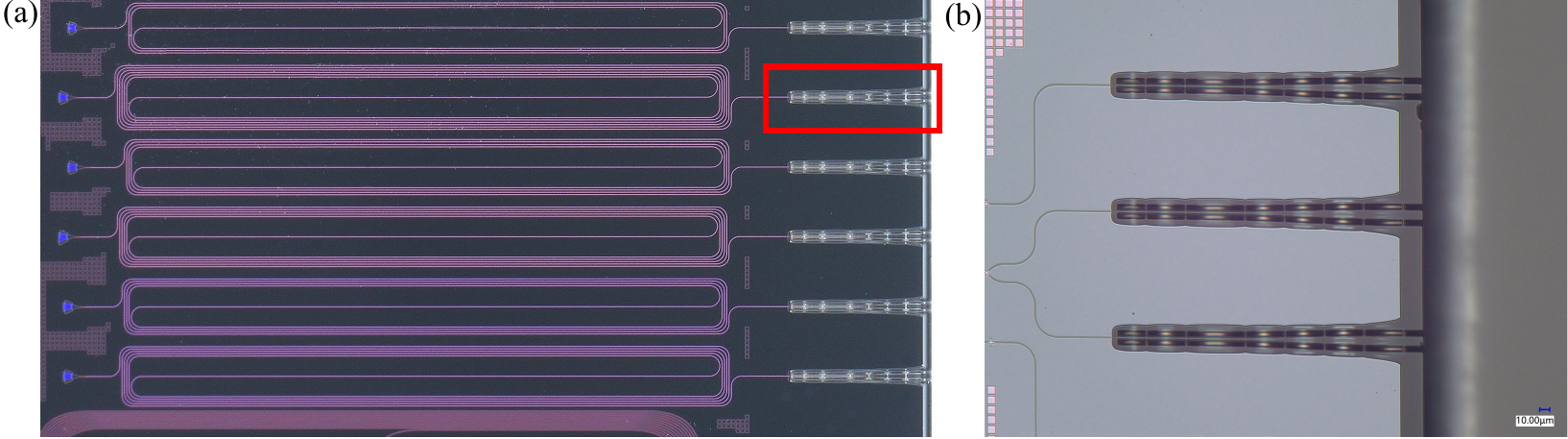}
	\caption{\label{F7}  
			\textbf{Photo of SOI chip.}
			(a) Spiral waveguides with different length.
			(b) Cantilever structure for chip-fiber coupling.
	}
\end{figure}

	The silicon chip (Fig.~\ref{F7}) we used for entanglement generation was fabricated through in-house 180~nm CMOS processes. Its waveguide structure was patterned using 248~nm deep ultraviolet (DUV) photolithography and inductively coupled plasma (ICP) etching processes.  
	The silicon-substrate-based wafer has a sub-BOX layer of 3~µm oxide, a core waveguide layer of 220~nm silicon, and a top cladding layer of 1~um oxide. 
	Fully etched single-mode waveguide with a dimension of $450\times220$~nm (width$\times$height) was used as the spontaneous four wave fixing (SFWM) source.
	We measure a propagation loss of 2.1~dB/cm for the $450\times220$~nm waveguide.
	
\begin{figure}[h]
\centering
\includegraphics[width=260px]{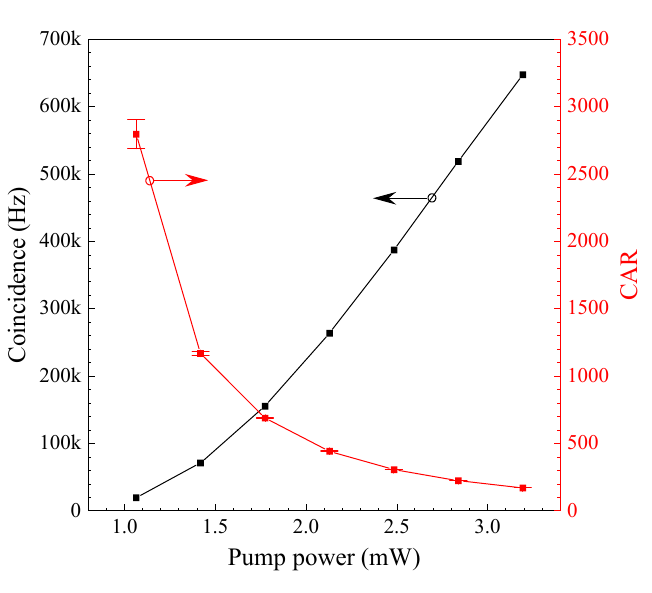}
\caption{\label{F8}  
		\textbf{Experimental coincidence rate and CAR of our source.}
		The black data points refer to the coincidence rates, and the red data points refer to the obtained CAR.
}
\end{figure}

	We characterize the performance of our entanglement source using photon-pair generation rate (PGR) and coincidence-to-accidental ratio (CAR).
	Figure~\ref{F8} plots the measured two figures of merit versus the on-chip pump power.
	To obtain the PGR, we implement polynomial fitting on the data of coincidence rate and have $\text{C.C.}=61719p^2+40290p-101779$, where C.C. denotes the coincidence rate and $p$ the pump power, in the unit of Hz.
	Combining with the collection efficiency of $-6.6$~dB (see main text), we derive that PGR$=1.3\times10^6$~pairs~s$^{-1}$~mW$^{-2}$.
	It is worth mentioning that the reported PGR does not present the best performance of our chip as we limited the pump power in order to obtain a modest coincidence-to-accidental ratio (CAR) of approximate 170.

	\newpage
\section*{Note 4: Phase stabilization of AMZI}
\begin{figure}[ht]
	\centering
	\includegraphics[width=500px]{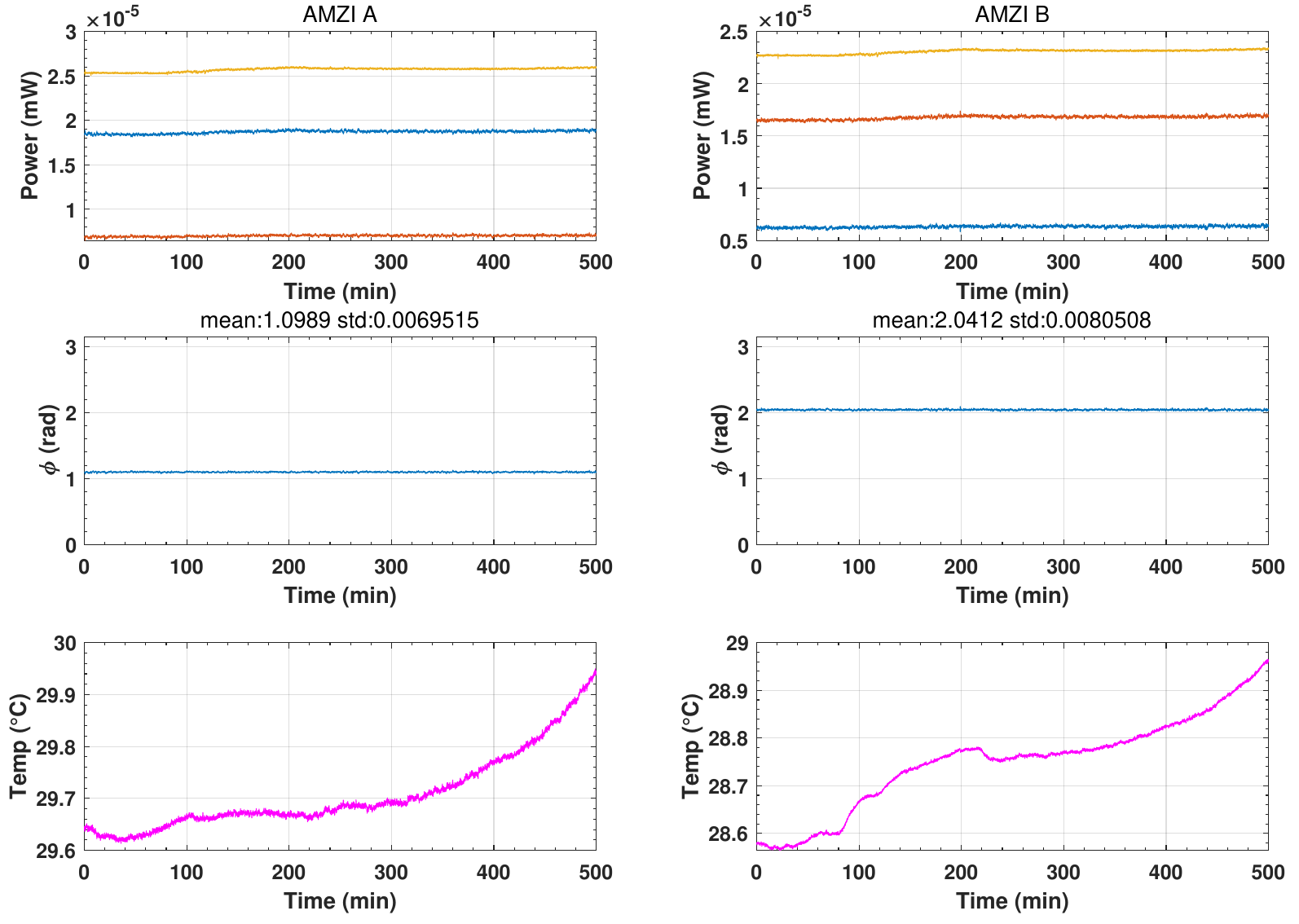}
	\caption{\label{F9}  
			\textbf{Phase log of AMZI.}
			The left column lists the data of Alice's AMZI, while the right column lists the data of Bob's AMZI.
			In the first row, the figure shows the output power of the AMZI of Alice and Bob, respectively.
			Specifically, the red line represents the power of output ``$+$", while the blue line represents the power of output ``$-$", and the yellow line represents the sum of output ``$+$" and ``$-$".
	}
\end{figure}

	We use two home-made asymmetric Mach-Zehnder interfrometers (AMZIs) in our quantum steering experiment.
    Each AMZI comprises two polarization-maintained 50:50 fiber couplers and a set of temperature control modules (including Thermoelectric Cooler (TEC), TEC driver, and thermistor), and is enclosed by a box to shield it from environmental temperature fluctuations.
	The differential delay of the AMZI is 400 ps, which is in line with the pump-pulse repetition frequency.
	The phase of each AMZI is actively stabilized using a continuous wave reference light centered at 1550.52~nm to provide the feedback control via temperature tuning.
	Our feedback system can stabilize the phase at an arbitrary value for over 500~minutes, see Fig.~\ref{F9}.

	\newpage
\section*{Note 5: Signal synchronization and phase calibration}

\begin{figure}[htp]
	\includegraphics[width=.6\columnwidth]{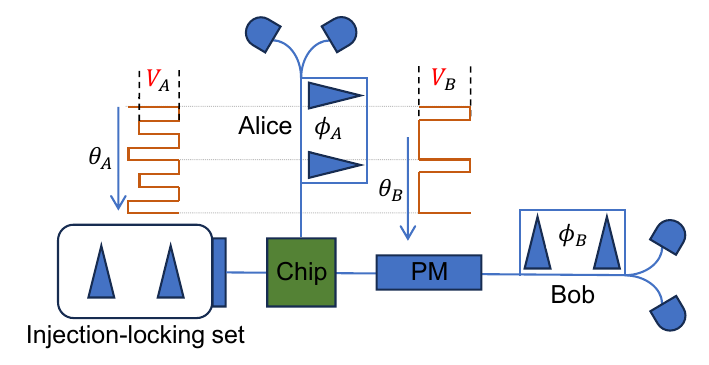}
	\caption{\label{F10} 
			\textbf{The schematic of phase calibration.}
			The master and slave lasers are combined as injection-locking set.}
\end{figure}

\textcolor{red}{
	We use the schematic shown in Fig.~\ref{F10} to illustrate the calibration procedure for the RF driving signals and the interferometer phases in the initial test runs. 
	Alice's injection-locking set outputs a train of laser pulses at a clock rate of 2.5~GHz, and these pulses are grouped in sequential pairs according to their differential phases. 
	Within each pair, the differential phase of the pulses is controlled to $\theta_A$ by adjusting the voltage levels in the RF driving signal to Alice's master laser~\cite{yuan2016Directly}. 
	Between different photon pairs, there is no fixed phase relationship because the master laser is gain-switched below the lasing threshold to assure phase randomization after each modulation clock (1.25~GHz). 
	Bob drives his phase modulator (PM) to impart a differential phase ($\theta_B$) to each signal pair before entering to his AMZI. 
	We use an arbitrary waveform generator (AWG) with internal delay lines to generate and synchronize the driving signals.} 

\textcolor{red}{For the purpose of phase calibration, we apply a pair of repetitive driving signal sequences to Alice's master laser and Bob's phase modulator.
} 
\textcolor{red}{
	We first fix $V_A=V_{A0}$ and impart a phase of $\theta_{A0}$ to the state. 
	Then we adjust the driving signal amplitude $V_B$ to the Bob's PM to maximize the correlation $S_n=|\max(S_n)|$ (perfect correlation). 
	In doing so, we achieve the following equation of phases according to Eq.~\eqref{phasecorrelation} 
	$$\theta_{A0}+\phi_A + \theta_{B0}+\phi_B =0 ,$$
	where $\phi_{A/B}$ denotes the relative phase induced by Alice's/Bob's AMZI.
    The synchronization of the RF driving signals are also achieved during this process through optimizing the relative delay of the two signals to ensure the certain $V_{B0}$ is as low as possible.}	
	
	\rev{Next, we increase $V_B$ to $V_{B1}$ to have $S_n=-|\max(S_n)|$ (perfect anti-correlation), in this way we achieve 
		$$\theta_{A0}+\phi_A + \theta_{B1}+\phi_B =\pi.$$
	Similarly, we fix $V_B=V_{B0}$ and increase $V_A$ to $V_{A1}$ to have $S_n=-|\max(S_n)|$, in this way we achieve 
		$$\theta_{A1}+\phi_A + \theta_{B0}+\phi_B =\pi.$$
	Now we have $\theta_{A1}-\theta_{A0}=\theta_{B1}-\theta_{B0}=\pi$, which corresponds to the voltage change of $V_{A1}-V_{A0}$ or $V_{B1}-V_{B0}$.
	That is to say, the voltage change is equivalent to a half-wave phase shift.
	We denote the above differences of voltage by $\Delta A$ and $\Delta B$, respectively.
	Given the phase calibration above, we divide the difference of voltage into $n-1$ segments and finally have the measurement $M^{(A)}_j$ corresponding to the voltage setting of $V_A=V_{A0}+\left(1-\frac{j-1}{n-1}\right)\Delta A$, and $M^{(B)}_j$ corresponding to the voltage setting of $V_B=V_{B0}+\frac{j-1}{n-1}\Delta B$, where $j \in \{1,...,n-1\}$.
	It is worth noting that the phase measurements are defined independent of the relative phases of the AMZIs.}

\newpage
\section*{Note 6: Calculation of $S_n$ and white noise assumption}
\subsection*{Calculation of $S_n$}

	For each pair of outputs of the AMZIs, there are three possible photon arrival times, which results in three equidistant coincidence peaks in the arrival-time histogram. 
	The coincident events in the side peaks yield detections in the time basis ($\hat\sigma_{\textsf{Z}}$).
	Those coincidence counts in the side peaks correspond to the pump photon annihilating in the early time-bin and the two daughter photons passing through the short arm of their respective AMZI, or correspond to the pump photon annihilating in the late time-bin and the two daughter photons passing through the long arm of their respective AMZI.
	We denote the coincidence counts in the side peaks by $CC_{time}$ and the ones in the adjacent time-bins by $CC_{acci}$.
	By remapping the measurement and outcome as we measure $-\hat\sigma_{\textsf{Z}}$ at Alice's site and $\hat\sigma_{\textsf{Z}}$ at Bob's site, we have $T=\langle A_0 \hat\sigma^B_0 \rangle=\frac{CC_{acci}-CC_{time}}{CC_{time}+CC_{acci}}$, which corresponds the term of $k=0$ in $S_n$.
	Note that $CC_{time}$ and $CC_{acci}$ comprises not only the counts of A$-$ \& $B-$ but all four combinations of outputs.

	On the other hand, the coincident events in the central peak yield detections in the phase basis ($\hat\sigma_{\theta}$), according to which we compute the rest of terms in $S_n$.
	Note that in the phase basis, the correlation is not about the detection times but between the detectors that count the photons.
	Those coincidence counts correspond to the superposition of the pump photon annihilating in the early/late time-bin and the two daughter photons passing through different arm of their respective AMZI.
	We denote the coincidence counts in the central peaks by $CC_{++}$, $CC_{+-}$, $CC_{-+}$, $CC_{--}$, respectively.
	When performing phase measurements, the terms of $k=1$ to $k=n-1$ in steering parameter can be obtained with $P=\frac{1}{n-1}\sum_{k=1}^{n-1}\langle A_k \hat\sigma^B_k \rangle
	=\frac{CC_{++}-CC_{+-}-CC_{-+}+CC_{--}}{CC_{++}+CC_{+-}+CC_{-+}+CC_{--}}$.
	Thus we have $S_n=\frac{1}{n}\left(T+(n-1)P\right)$.
	
\subsection*{The isotropic noise assumption}
	We justify our assumption on the form of noise for the following reasons.
	(i) It is a common treatment in experiments using other degrees of freedom, such as polarization~\cite{saunders2010Experimental} and orbital angular momentum~\cite{zeng2018}, where the deliberately introduced white noises are arguably unbiased to the measurement settings.  
	(ii) Among all devices in our setup, including the DFB laser, phase modulator, fiber, SOI chip etc., no apparent phase dependence is found, which means the phase fluctuation induced by the devices should be uniformly distributed ranging from 0 to $2\pi$ and thus unbiased to the phase measurement we perform.
	In terms of experimental results, by performing a pair of fixed complementary measurement on the two parties, such as $\hat\sigma_0$ and $\hat\sigma_\pi$, we find that the erroneous coincidence counts are arguably the same as the ones with other pairs of complementary measurements, which supports our statement.
	(iii) Regarding the noise in the time basis, as we can see from Fig.~4\textbf{b} in the main text, the accidental counts are arguably uniformly distributed in the time-bins, which shows the unbiasedness of the noise to time basis.
	(iv) Moreover, the contrast of correlated counts and erroneous counts is at the same level to the contrast in the phase basis, which means the noise is unbiased between in time and phase basis.

\section*{References}

\bibliography{Refs.bib}